# Exotic spin, charge and pairing correlations of the two-dimensional doped Hubbard model: a symmetry entangled mean-field approach

O. Juillet[1] and R. Frésard[2]

*[1] Laboratoire LPC Caen, ENSICAEN, Université de Caen, CNRS/IN2P3,
6 Boulevard Maréchal Juin, 14050 Caen Cedex, France*

*[2] Laboratoire CRISMAT, UMR CNRS-ENSICAEN 6508,
6 Boulevard Maréchal Juin, 14050 Caen Cedex, France*

Intertwining of spin, charge and pairing correlations in the repulsive two-dimensional Hubbard model is shown through unrestricted variational calculations, with projected wavefunctions free of symmetry breaking. A crossover from incommensurate antiferromagnetism to stripe order naturally emerges in the hole-doped region when increasing the on-site coupling. Although effective pairing interactions are identified, they are strongly fragmented in several modes including *d*-wave pairing and more exotic channels related to an underlying stripe. We demonstrate that the entanglement of a mean-field wavefunction by symmetry restoration can largely account for interaction effects.

Transition metal oxides are prototypical strongly correlated systems that exhibit a rich variety of quantum phenomena such as antiferromagnetism (AF), incommensurate charge and spin ordering or high-$T_c$ superconductivity. The understanding of their subtle competition at low-temperature remains one of the most challenging topics in condensed matter physics. According to Anderson's proposal [1], the two-dimensional (2D) single-band Hubbard model is expected to provide a minimal framework for addressing these issues in rare earth cuprates. It describes *d*-electrons hopping between the Wannier states of neighboring lattice sites in a copper-oxygen plane experiencing a purely local Coulomb repulsion. At half-filling, the interaction strength drives a Mott transition to an insulator [2] with long-range AF order [3]. When the lattice is doped away from half-filling, the exact form of the phase diagram is still controversial. Of central interest, in this regime, is whether the ground state supports unconventional fermion-pair condensates or charge inhomogeneities, and if so, how their order parameters are intertwined with magnetic properties. Only a partial answer can be currently obtained through approximate many-body techniques, such as cluster extensions [4,9] of the dynamical mean-field theory [5,9], the two-particle self-consistent approximation [6,9], Gutzwiller variational schemes [7,9] or slave-boson approaches [8,9]. Standard quantum Monte-Carlo simulations (QMC) are also restricted [10] owing to the notorious sign problem that is particularly severe for doped Hubbard models. Although new sign-free stochastic reformulations have been recently introduced [11,12], they are not, for now, immune to systematic errors [13]. To overcome these theoretical difficulties, a direct quantum simulation has been suggested by loading ultracold mixtures of two interacting fermionic species in optical lattices. Indeed, such atomic systems allow for an almost perfect implementation of the Hubbard model with tunable parameters [14]. The crossover from a metallic into a Mott insulating regime has already been observed [15], while magnetic ordering and potential exotic superfluidity remains to be achieved using, e.g., new cooling techniques [16].

Most variational investigations of low-lying states in the Hubbard model are carried out starting from a simple mean-field wave-function to include quantum correlation effects through projection techniques. For instance, the celebrated Gutzwiller ansatz suppresses (totally or partially) double occupancy in the strong correlation regime from a *d*-wave superconducting state [17] or from a Slater determinant with assumed magnetic and charge orders [17,18]. Such a procedure substitutes for an unrestricted calculation, which is currently numerically untractable but highly desirable to identify the low-energy orders that spontaneously emerge from the 2D Hubbard model. The purpose of this letter is to extract spin, charge and pairing correlations that result from an unbiased energy minimization using an alternative projected wave-function.

The single-band Hubbard Hamiltonian on a *D*-dimensional lattice with periodic boundary conditions may be written as

$$\hat{H} = -t \sum_{\langle \vec{r},\vec{r}'\rangle \sigma} \hat{c}^+_{\vec{r}\sigma}\hat{c}_{\vec{r}'\sigma} + U \sum_{\vec{r}} \hat{n}_{\vec{r}\uparrow}\hat{n}_{\vec{r}\downarrow} \qquad (1)$$

where $t$ is the hopping integral between neighboring sites $\langle \vec{r},\vec{r}'\rangle$ and $U$ the repulsive local Coulomb interaction; $\hat{c}^+_{\vec{r}\sigma}$, $\hat{c}_{\vec{r}\sigma}$ and $\hat{n}_{\vec{r}\sigma} = \hat{c}^+_{\vec{r}\sigma}\hat{c}_{\vec{r}\sigma}$ are, respectively, electronic creation, annihilation and density operators at site $\vec{r}$ with spin label $\sigma \in \{\uparrow,\downarrow\}$. To describe the lowest energy *N*-electron eigenstate for given quantum numbers $\Gamma$, the variational approach we propose relies on a trial Slater determinant $|\Phi\rangle = \hat{c}^+_{\phi_1}\cdots\hat{c}^+_{\phi_N}|0\rangle$ projected on the symmetry subspace associated to $\Gamma$. Here, $\hat{c}^+_{\phi_n} = \sum_{\vec{r}\sigma} \hat{c}^+_{\vec{r}\sigma}\phi_n(\vec{r}\sigma)$ creates one electron in the spinor wave-function $\phi_n(\vec{r}\sigma)$. In this way, the variational ansatz is no longer a single Slater determinant but a linear superposition of product states that can absorb electronic correlations beyond mean-field. Apart from orthogonality and normalization constraints $\langle\phi_n|\phi_{n'}\rangle = \delta_{n,n'}$ on the single-particle states, no restriction is imposed to the mean-field state $|\Phi\rangle$. In particular, the usual factorization of $|\Phi\rangle$ into a product of up-spin and down-spin Slater determinants has turned out to be inefficient. In all following numerical applications, we allow mono-electronic states to develop spin textures in the *x-y* plane. For the Hubbard model (1), $\Gamma$ quantum numbers include total pseudo-momentum $\hbar\vec{K}$, total spin $S$ and its *z*-component $S_z$, as well as labels associated to discrete lattice symmetries. In the language of group theory, $\Gamma$ defines an irreducible representation of the Hamiltonian symmetry group. A symmetry-adapted mean-field state can thus be obtained by applying the standard projection operator given by [19]:

$$\hat{P}^{(\Gamma)} = \frac{d^{(\Gamma)}}{\Omega} \sum_g \left(\chi_g^{(\Gamma)}\right)^* \hat{U}_g \qquad (2)$$

where the sum runs over all symmetry transformations $g$ realized by unitary operators $\hat{U}_g$ in the many-body Hilbert space. The set $\{\chi_g^{(\Gamma)}\}$ corresponds to the character of the representation $\Gamma$ and $d^{(\Gamma)}$ to its dimension. $\Omega = \sum_g 1$ is the order of the symmetry group. In the case of continuous transformations, the sum has to be supplemented by group integration with the Haar mesure [17]. For instance, $SU(2)$ spin-rotational symmetry restoration can be performed using Euler's angles $(\alpha,\beta,\gamma)$ parametrization of rotations, i.e $\hat{U}_g = \exp(-i\alpha \hat{S}_z/\hbar)\exp(-i\beta \hat{S}_y/\hbar)\exp(-i\gamma \hat{S}_z/\hbar)$ with $\hat{\vec{S}}$ the total spin observable. The projection property $\left(\hat{P}^{(\Gamma)}\right)^2 = \hat{P}^{(\Gamma)}$ and the Hamiltonian invariance under symmetry group transformations $\hat{U}_g$ allow to cast the average energy $\mathcal{E}^{(\Gamma)}$ in the projected state $\hat{P}^{(\Gamma)}|\Phi\rangle$ in the form

$$\mathcal{E}^{(\Gamma)} = \frac{\langle\Phi|\hat{H}\hat{P}^{(\Gamma)}|\Phi\rangle}{\langle\Phi|\hat{P}^{(\Gamma)}|\Phi\rangle} \qquad (3)$$

From Wick's theorem, the expectation values of the many-body operators $\hat{H}\hat{P}^{(\Gamma)}$ and $\hat{P}^{(\Gamma)}$ in the Slater determinant $|\Phi\rangle$ are only expressed in terms of the contractions $\rho_{\vec{r}\sigma,\vec{r}'\sigma'} = \langle\hat{c}^+_{\vec{r}'\sigma'}\hat{c}_{\vec{r}\sigma}\rangle_\Phi = \sum_n \phi_n(\vec{r}\sigma)\phi_n^*(\vec{r}'\sigma')$. Therefore, the projected energy (3) is a functional of $\rho$: $\mathcal{E}^{(\Gamma)} = \mathcal{E}^{(\Gamma)}[\rho]$. By introducing a $N \times N$ hermitian matrix of Lagrange multipliers $\lambda_{nn'}$, stationarity of the Lagrangian function $\mathcal{E}^{(\Gamma)} - \sum_{n,n'}\lambda_{nn'}\langle\phi_n|\phi_{n'}\rangle$ with respect to single-particle states immediately leads to the self consistency equation $\left[h^{(\Gamma)},\rho\right] = 0$, where $h^{(\Gamma)}$ is an effective single-particle Hamiltonian defined by the following matrix elements:

$$h^{(\Gamma)}_{\vec{r}\sigma,\vec{r}'\sigma'} = \frac{\partial \mathcal{E}^{(\Gamma)}}{\partial \rho_{\vec{r}'\sigma',\vec{r}\sigma}} \qquad (4)$$

These results remain obviously valid irrespective of the precise form of the operator $\hat{P}^{(\Gamma)}$. In the mean-field approach, $\hat{P}^{(\Gamma)}$ is set to identity and one recovers from Eq. (3) the usual energy functional $\mathcal{E}[\rho]$ and from Eq. (4) the associated single-particle Hartree-Fock Hamiltonian:

$$h[\rho] = -t\sum_{\langle\vec{r},\vec{r}'\rangle\sigma}|\vec{r}\sigma\rangle\langle\vec{r}'\sigma| + U\sum_{\vec{r}\sigma}\left(\rho_{\vec{r}\bar{\sigma},\vec{r}\bar{\sigma}}|\vec{r}\sigma\rangle\langle\vec{r}\sigma| - \rho_{\vec{r}\sigma,\vec{r}\bar{\sigma}}|\vec{r}\sigma\rangle\langle\vec{r}\bar{\sigma}|\right) \qquad (5)$$

where $\sigma$ and $\bar{\sigma}$ are time-reversed conjugate spin states. Consequently, with or without symmetry restoration, optimal spin-orbitals $\phi_n(\vec{r}\sigma)$ are obtained as eigenvectors of the state-dependent Hamiltonian (4). A similar conclusion has been drawn previously in nuclear physics [20] and for molecular electronic structure [21]. The projected-energy derivative with respect to $\rho$ can be further calculated by noting that the transformed state $|\Phi_g\rangle = \hat{U}_g|\Phi\rangle$ is again a Slater determinant. Owing to extended Wick's theorem for matrix elements [22], the symmetry entangled mean-field (SEMF) Hamiltonian (4) finally becomes

$$h^{(\Gamma)}[\rho] = \frac{1}{\sum_g \chi_g^{(\Gamma)} \det A_g}\sum_g \chi_g^{(\Gamma)} A_g^{-1}\left\{\left(U_g - 1\right)\left(\mathcal{E}\left[\mathcal{R}_g\right] - \mathcal{E}^{(\Gamma)}\right) + h\left[\mathcal{R}_g\right]U_g B_g^{-1}\right\} \qquad (6)$$

Here the matrix $A_g$, $B_g$ depend on the one-body density $\rho$ and on the matrix representation $U_g$ in the one-electron space of group transformations, according to:

$$A_g = 1 + \left(U_g - 1\right)\rho, \quad B_g = 1 + \rho\left(U_g - 1\right) \qquad (7)$$

$\mathcal{R}_g = \rho U_g A_g^{-1}$ is the transition one-body density matrix [22] between the uncorrelated state $|\Phi\rangle$ and its symmetry transformed counterpart $|\Phi_g\rangle$: $\left(\mathcal{R}_g\right)_{\vec{r}\sigma,\vec{r}'\sigma'} = \langle\Phi|\hat{c}^+_{\vec{r}'\sigma'}\hat{c}_{\vec{r}\sigma}|\Phi_g\rangle / \langle\Phi|\Phi_g\rangle$. In practice, we restore all symmetries of the Hamiltonian (1) that lead, for the rectangular $16 \times 4$ cell we consider below, to a variational wave-function formed by the linear superposition of about $10^6$ symmetry related Slater determinants. On small $4 \times 4$ clusters, where Hamiltonian diagonalization can be fully performed, we show in the supplemental material that the ground-state energy and various correlation functions differ respectively by less than 1% and

2.5% with respect to exact results. At half-filling, where unbiased QMC calculations have been performed [10], the relative error on the energy at $U = 4t$ is 0.5% (1.5%) on the $6 \times 6$ ($8 \times 8$) lattice. In several cases, the SEMF approach outperforms the usual Gutzwiller variational scheme.

Let us now focus on magnetic and charge correlations in the hole doped region from intermediate to strong coupling. Note that large cell sizes are required for the development of possible incommensurate ordering with long wavelength. To meet this challenge by means of a variational calculation with an unrestricted projected wave-function, we applied the above symmetry entangled mean-field approach limiting here ourselves to a rectangular $16 \times 4$ cell, at doping $\delta = 1/8$ with periodic boundary conditions. A smaller dimension in the $x$-direction would not allow to capture translationally invariant filled vertical stripe phases that are expected from constrained-path QMC calculations [23]. Only zero total pseudo-momentum and spin singlet states are addressed. These quantum numbers rigorously characterize the ground-state at half-filling on a square lattice [24,25]. They are also found for the hole-doped case in exact diagonalization [26] and path-integral renormalization group studies [27] on small clusters. Furthermore, Nagaoka ferromagnetism, which could invalidate such a statement, is not expected for the relatively moderates values $U \leq 12t$ of the on-site repulsion we consider. Finally, the $A_1$ irreducible representation of the $C_{2v}$ lattice group, that embraces $s$-wave and $d$-wave symmetries, is imposed. Since our variational state is translationally and spin rotationally invariant, possible magnetic and charge ordering can only be highlighted through correlation functions or their Fourier transform. Hence, we calculate the spin $S_m(\vec{k})$ and charge $S_c(\vec{k})$ structure factors defined by:

$$S_m(\vec{k}) = \frac{4}{3} \sum_{\vec{r}} \exp(i\vec{k}\vec{r}) \left\langle \hat{\vec{S}}_{\vec{0}} \hat{\vec{S}}_{\vec{r}} \right\rangle_{P^{(\Gamma)}\Phi}, \quad S_c(\vec{k}) = \sum_{\vec{r}} \exp(i\vec{k}\vec{r}) \left\langle \delta\hat{n}_{\vec{0}} \delta\hat{n}_{\vec{r}} \right\rangle_{P^{(\Gamma)}\Phi} \quad (8)$$

where $\hat{\vec{S}}_{\vec{r}} = \frac{1}{2} \sum_{\sigma,\sigma'} \hat{c}^+_{\vec{r}\sigma} \vec{\tau}_{\sigma,\sigma'} \hat{c}_{\vec{r}\sigma'}$ is the spin operator at lattice node $\vec{r}$ (with $\vec{\tau}$ the usual Pauli matrices) and $\delta\hat{n}_{\vec{r}} = \sum_{\sigma} \left( \hat{n}_{\vec{r}\sigma} - \left\langle \hat{n}_{\vec{r}\sigma} \right\rangle_{P^{(\Gamma)}\Phi} \right)$ the local charge fluctuation. Figure 1.a shows $S_m(\vec{k})$ in the upper right quarter of the first Brillouin zone for relevant parameter values. It entails two main features. First, a weakly $\vec{k}$-dependent broad background manifestly appears and is almost insensitive to the interaction. Second, with increasing $U/t$, the peak at the AF wavevector $\vec{k} = (\pi, \pi)$ decreases (to the extend that it vanishes for $U \geq 8t$) and it leaves place to incommensurate magnetic correlations signaled by the symmetry related peaks at $\vec{k} = (\pm 7\pi/8, \pi)$. This physically corresponds to staggered magnetization periodically modulated with the wavelength $\lambda_m = 16 = 2/\delta$. The crossover that may be inferred from these spin-spin correlations is further confirmed by the charge structure factor $S_c(\vec{k})$ shown in Fig. 1b. Indeed, for the intermediate coupling $U = 4t$, no particular signal emerges. On the contrary, from $U = 6t$ to $U = 12t$, a peak at $\vec{k} = (\pm \pi/4, 0)$ gradually develops. It follows from the translationally invariant superposition of broken symmetry charge-density wave states associated to a hole density profile of wavelength $\lambda_c = 8$. Since $\lambda_c = \lambda_m/2$, this is the signature of filled vertical stripes at the boundaries of AF domains separated by $\pi$-phase shifts, in agreement with the solitonic mechanism proposed by Zaanen and Oles [28]. In short, we have observed, with SEMF wave-functions, the crossover predicted by constrained-path

QMC [23] from a spin-density wave to a stripe-like state with increasing Coulomb interaction.

We now address the development of pairing correlations according to Yang's criteria for off-diagonal long-range order [29]. Upon the diagonalization of a pair-field correlation matrix $\mathcal{P}_{\vec{r},\vec{r}'}$, superconductivity appears as long as one or more eigenvalues scale as the particle number when the system size is increased at fixed density. The associated eigenvectors define pair wavefunctions. Here, we deal with a finite system and only electron pairing modes favored by the interaction may be identified. Moreover, one has to discard fictitious correlations from non-interacting dressed electrons and that only vanish at the thermodynamical limit. Hence, for spin-singlet pairing, we use the following matrix $\mathcal{P}_{\vec{r},\vec{r}'}$:

$$\mathcal{P}_{\vec{r},\vec{r}'} = \frac{1}{2}\left\langle \hat{\Pi}_{\vec{r}}^+ \hat{\Pi}_{\vec{r}'} + \hat{\Pi}_{\vec{r}} \hat{\Pi}_{\vec{r}'}^+ \right\rangle_{P^{(\Gamma)}\Phi}^{(V.C.)} \qquad (9)$$

where $\hat{\Pi}_{\vec{r}}^+ = \frac{1}{\sqrt{2\mathcal{N}}} \sum_{\vec{R}} \left( \hat{c}_{\vec{R}\uparrow}^+ \hat{c}_{\vec{R}+\vec{r}\downarrow}^+ - \hat{c}_{\vec{R}\downarrow}^+ \hat{c}_{\vec{R}+\vec{r}\uparrow}^+ \right)$ is the translationally invariant and spin singlet pair-field operator for two electrons separated by $\vec{r}$ and $\mathcal{N}$ the number of lattice sites; $\langle\cdots\rangle^{(V.C.)}$ stands for the vertex contribution [30]. We display on Table 1 the first eigenvalues of the $\mathcal{P}$-matrix in the strongly repulsive regime $U = 12t$ as compared to those obtained for the attractive coupling $U = -4t$. In both cases, positive eigenvalues are obtained. They purely result from the entanglement realized by symmetry restoration: Without projection, the matrix elements $\mathcal{P}_{\vec{r},\vec{r}'}$ are all vanishing. For the 2D attractive Hubbard model, off-diagonal long-ranged order associated to BCS $s$-wave superconductivity is well established for moderate interactions, especially in the vicinity of quarter filing [31]. The SEMF approach gives here good evidence for a precursor to the superconducting behavior, since one pairing eigenmode clearly outperforms the other ones. Truly, the associated pair wavefunction exhibits a large on-site component accompanied with a small extended-$s$ contribution. In the repulsive regime, perhaps the most striking result consists in highlighting the emergence of an effective attractive interaction in the particle-particle channel. However, it is strongly fragmented as all first few eigenvalues are nearly degenerate. Concerning the pairing symmetry, Fig. 2 presents the wavefunction associated to the leading modes. On top of the expected $d$-wave (Fig. 2.a) and extended $s$-wave (Fig. 2.e) pairing [32], exotic competitive modes related to stripe-like correlations are found. They are characterized by pairing at distances close to the charge wavelength $\lambda_c = 8$. Qualitatively, the formation of a singlet pair in a correlated AF background can be enhanced at this distance $\lambda_c = \lambda_m/2$ provided a domain wall separates the two electrons. This is schematically depicted in Fig. 3. No significant deviation appears when reducing $U$ in the stripy regime. For the intermediate coupling $U = 4t$, where no charge order can be detected, only $d$-wave (Fig. 2.a) and extended $s$-wave pairings are dominant.

In conclusion, we have performed unrestricted variational calculations for the 2D hole-doped Hubbard model in a theoretical framework where electronic correlations are generated by the entanglement of a mean-field wavefunction through the restoration of all Hamiltonian symmetries. From moderate to strong coupling, independent energy minimizations lead to consistent spin, charge and pairing correlations at doping $\delta = 1/8$. The spontaneous appearance of incommensurate spin-density waves evolving into striped states is accompanied by the development of several competing pairing channels. It remains to be elucidated if one of them would dominate when enlarging the variational subspace to bring out a precursor signal to unconventional superconductivity. Work along this line is in progress.

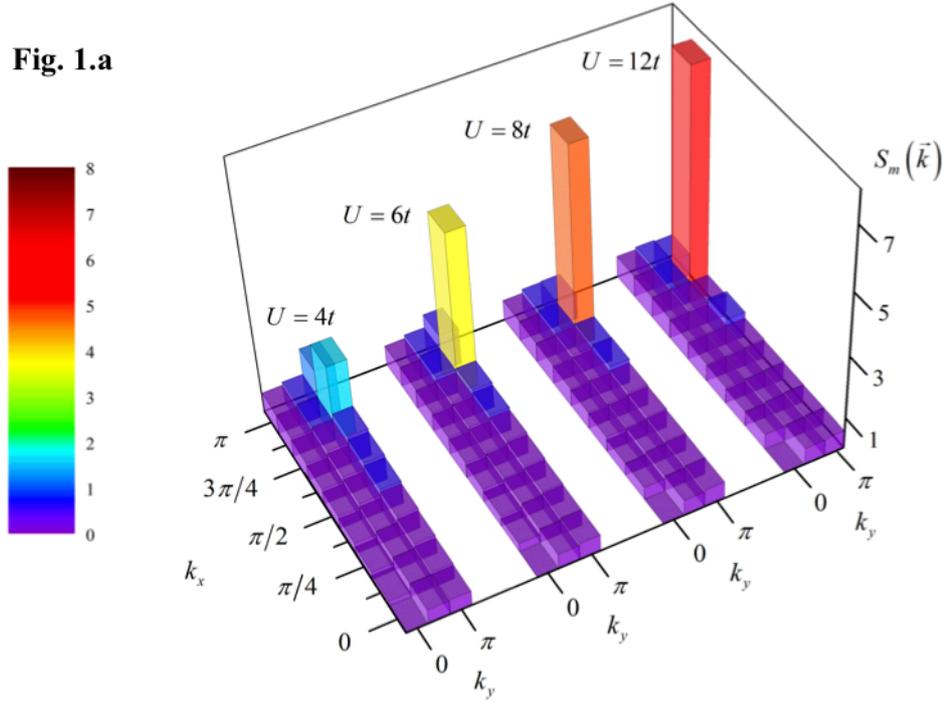

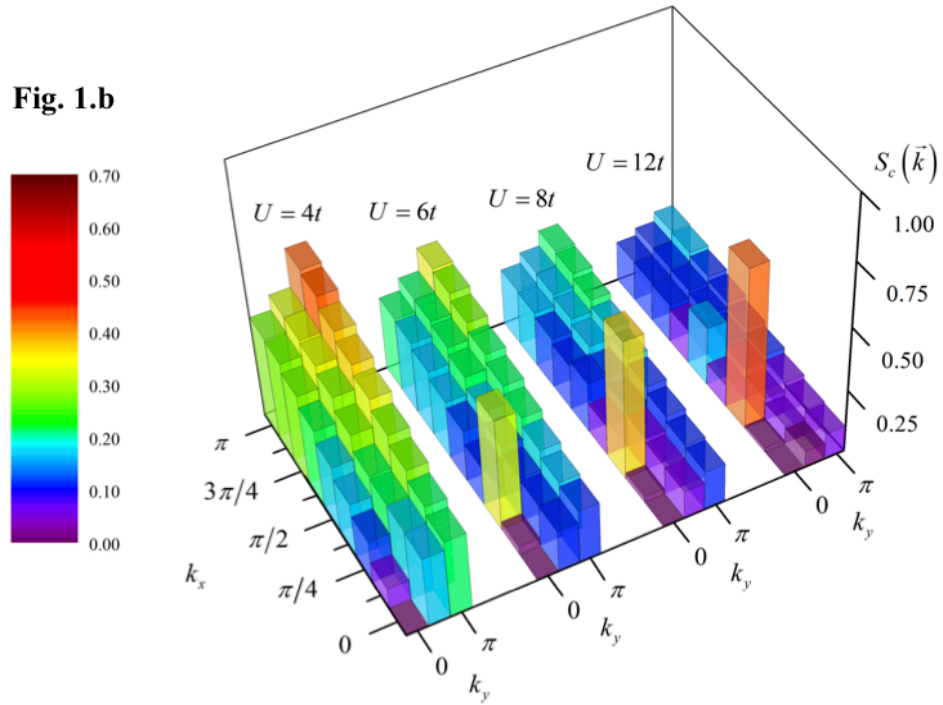

**FIG. 1 (color online)**. Momentum dependence of the magnetic (a) and charge (b) structure factors for a rectangular cell $16 \times 4$ at the hole doping $\delta = 1/8$ and for several values of the interaction strength $U/t$

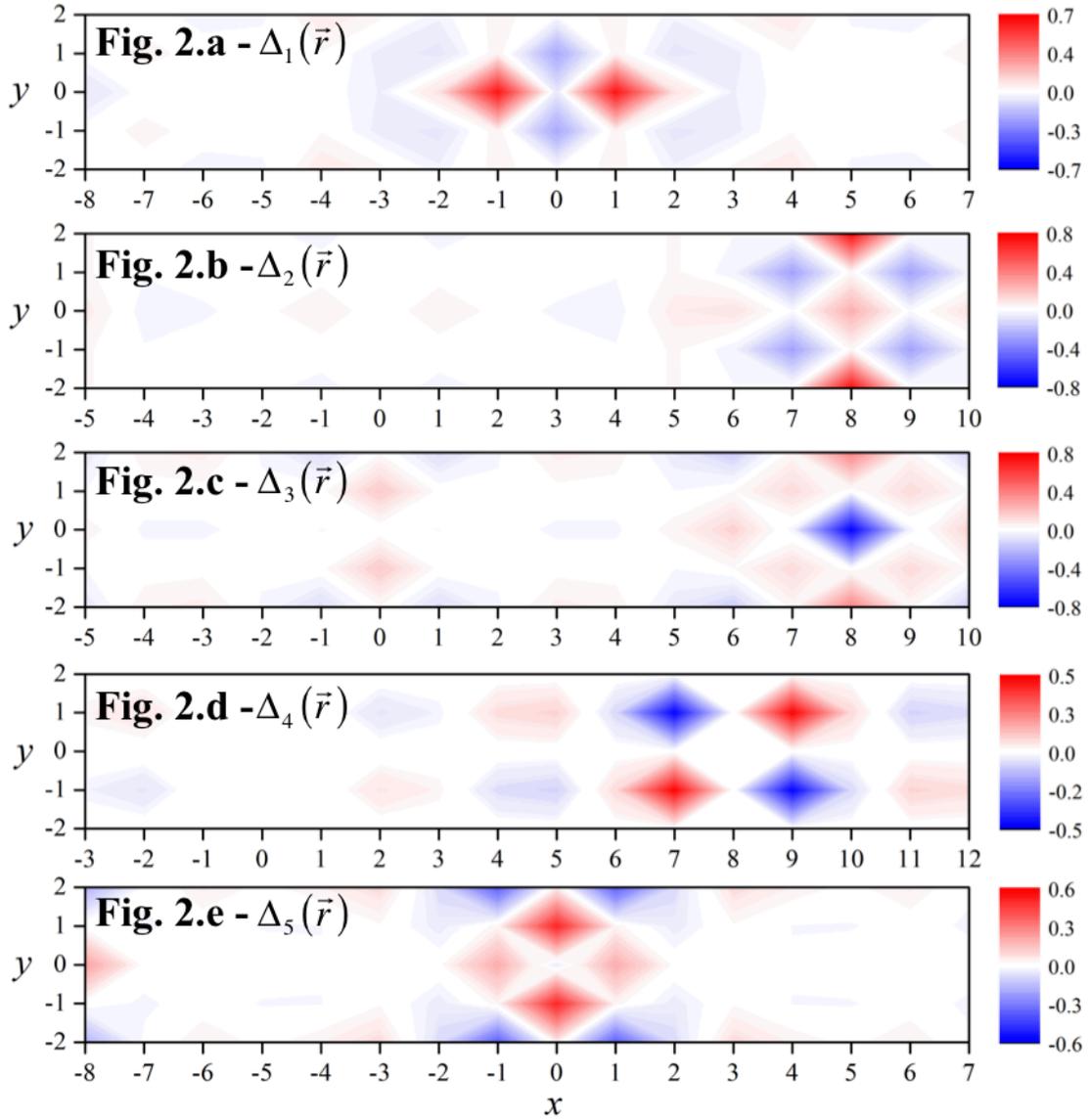

**FIG. 2 (color online).** Wavefunctions $\Delta_i(\vec{r})$ associated to the pairing modes defined by the eigenvectors of the $\mathcal{P}$-matrix (9) with the largest eigenvalues $\lambda_i$ shown on Fig. 2. $\hat{\Delta}_i^+ = \frac{1}{\sqrt{2\mathcal{N}}} \sum_{\vec{r},\vec{R}} \Delta_i(\vec{r}) \left( \hat{c}_{\vec{R}\uparrow}^+ \hat{c}_{\vec{R}+\vec{r}\downarrow}^+ - \hat{c}_{\vec{R}\downarrow}^+ \hat{c}_{\vec{R}+\vec{r}\uparrow}^+ \right)$ is the corresponding pair-field operator. A lattice $16 \times 4$ with $U = 12t$ is considered at hole doping $\delta = 1/8$.

## Fig. 3.a

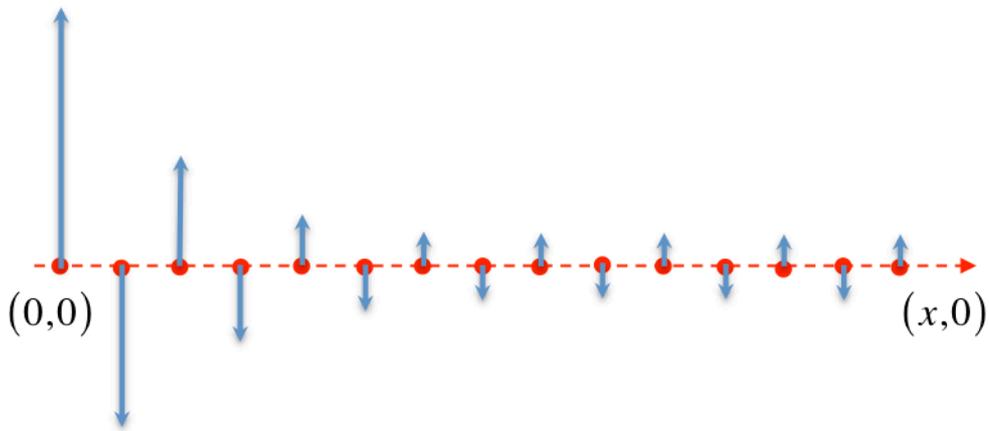

## Fig. 3.b

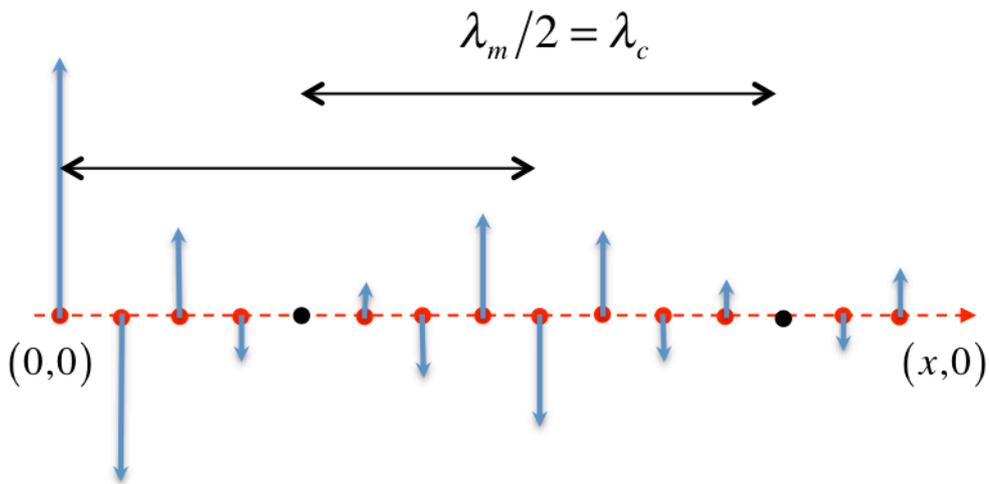

**FIG. 3 (color online).** Schematic view of singlet electron pairing in a correlated AF background (a) and in presence of a domain wall associated to an underlying filled stripe (b). (a) The length of the individual spins mimics the decrease of the spin autocorrelation function at small distances together with its long-ranged constant staggered magnetization. Thus, only nearest neighbor pairing dominates. (b) With holes localized in a vertical filled stripe, the incommensurate magnetic peak implies the repetition of a similar AF spin pattern between domain walls, with a $\pi$ phase-shift at the nodes. Therefore, the formation of singlet pairs at distances $x = \lambda_m/2 (= \lambda_c)$ is favored.

| $U/t$ | $N$ | $\lambda_1$ | $\lambda_2$ | $\lambda_3$ | $\lambda_4$ | $\lambda_5$ |
|---|---|---|---|---|---|---|
| −4 | 30 | 0.421 | 0.051 | 0.046 | 0.030 | 0.020 |
| 12 | 56 | 0.495 | 0.455 | 0.442 | 0.431 | 0.425 |

**TABLE I.** First eigenvalues $\lambda_i$ of the pairing correlation matrix $\mathcal{P}_{\vec{r},\vec{r}'}$ defined by Eq. (9). Results are shown for a cluster $16\times 4$ in an *s*-wave BCS-like superconducting regime ($N = 30$ electrons, $U = -4t$) and in the striped state obtained for the strong Coulomb repulsion $U = 12t$ at hole doping $\delta = 1/8$ ($N = 56$ electrons).

# Support material for "Exotic spin, charge and pairing correlations of the two-dimensional doped Hubbard model: a symmetry entangled mean-field approach"

O. Juillet[1] and R. Frésard[2]

[1] *Laboratoire LPC Caen, ENSICAEN, Université de Caen, CNRS/IN2P3, 6 Boulevard Maréchal Juin, 14050 Caen Cedex, France*

[2] *Laboratoire CRISMAT, UMR CNRS-ENSICAEN 6508, 6 Boulevard Maréchal Juin, 14050 Caen Cedex, France*

We summarize here the comparison of results computed with the symmetry entangled meanfield (SEMF) approach to existing data obtained by unbiased numerical simulations, i.e. exact diagonalizations and quantum Monte-Carlo (QMC) calculations at half-filling or in the attractive regime. Energies are reported in Table I and we focus in Table II on the magnetic $S_m(\vec{k})$ and charge $S_c(\vec{k})$ structure factors at the antiferromagnetic (AF) wavevector $\vec{k}=(\pi,\pi)$. On the $4\times 4$ lattice, the discrepancy does not exceed 1% for the energies and 2.5% for the considered correlation functions. As the lattice size is increased, the SEMF approach exhibits larger departures from QMC results for half-filled clusters and, in particular, overestimates the staggered magnetization. However, a similar behavior is also observed in variational Monte-Carlo (VMC) calculations with Gutzwiller projection on top of meanfield wavefunctions with assumed broken symmetries [1,2]. Indeed, the present unrestricted method gives approximate ground-state expectation values comparable or in some cases superior to the Gutzwiller scheme. For instance, on a $8\times 8$ half-filled cluster with periodic-antiperiodic boundary conditions at intermediate coupling $U=8t$, the SEMF variational energy per site is $-0.511t$ while the AF wavefunction with a Gutzwiller projection parameter gives $-0.493(3)t$ [2]. The staggered magnetization $M$, estimated from the spin structure factor according to $M=\sqrt{S_m(\pi,\pi)/\mathcal{N}}$ (with $\mathcal{N}$ the number of sites), is $M=0.501$ as compared to $M=0.86(1)$ for the Gutzwiller ansatz. In the hole-doped region, QMC methods suffer from severe sign problems that prohibit accessing low temperature and large clusters. Therefore, the VMC method remains a milestone. For 4 holes on a $6\times 6$ cluster (with periodic-antiperiodic boundary conditions), several Gutzwiller projected phases have been considered in [3]. The best energy, $-20.86(6)t$, is obtained with incommensurate AF and BCS $d$-wave pairing. The variational SEMF approach yields the lower value $-21.227t$.

Finally, note that a direct energy minimization with a symmetry projected Slater determinant has already been tested in one dimension [4] and very recently on small 2D clusters [5]. However, the formulation in terms of the SEMF Hamiltonian, defined by Eq. (6) in the main text and that plays a crucial role in our simulations for larger systems, was not considered. In addition, no lattice symmetry was restored.

| Lattice | $U/t$ | $t'/t$ | $N$ | $S$ | $\vec{K}$ | $\Gamma_{C_{4v}}$ | $E/t$ SEMF | $E/t$ Exact |
|---|---|---|---|---|---|---|---|---|
| $4\times 4$ | 4 | 0 | 10 | 0 | (0,0) | $A_1$ | −19.573 | −19.58 |
| $4\times 4$ | 8 | 0 | 10 | 0 | (0,0) | $A_1$ | −17.457 | −17.51 |
| $4\times 4$ | −4 | 0 | 10 | 0 | (0,0) | $A_1$ | −32.642 | −32.733 |
| $4\times 4$ | 4 | 0 | 14 | 0 | (0,0) | $B_1$ | −15.718 | −15.744 |
| $4\times 4$ | 8 | 0 | 14 | 0 | (0,0) | $B_1$ | −11.783 | −11.868 |
| $4\times 4$ | 12 | 0 | 14 | 0 | (0,0) | $A_1$ | −9.917 | −10.048 |
| $4\times 4$ | 8 | −0.3 | 14 | 0 | (0,0) | $A_1$ | −12.356 | −12.502 |
| $4\times 4$ | 4 | 0 | 16 | 0 | (0,0) | $A_1$ | −13.616 | −13.622 |
| $4\times 4$ | 8 | −0.3 | 16 | 0 | (0,0) | $A_1$ | −8.485 | −8.488 |
| $6\times 6$ | 4 | 0 | 36 | 0 | (0,0) | $B_1$ | −30.733 | −30.87(2) |
| $6\times 6$ | −2 | 0 | 18 | 0 | (0,0) | $A_1$ | −53.434 | −53.6(3) |
| $8\times 8$ | 4 | 0 | 64 | 0 | (0,0) | $A_1$ | −54.278 | −55.09(6) |

**TABLE I.** Comparison of Hubbard model ground-state energies $E$ from SEMF simulations with exact results. Periodic boundary conditions are imposed on both directions. $U$ is the on-site Coulomb repulsion and $t'$ corresponds to the hopping integral between next-nearest neighbors ; $N$ is the number of electrons, $S$ the total spin and $\hbar\vec{K}$ the total momentum; $\Gamma_{C_{4v}}$ denotes the irreducible representation of the square lattice symmetry group (an $s$−wave corresponds to the $A_1$ representation and a $d_{x^2-y^2}$−wave transforms according to the representation $B_1$). Exact diagonalization results are taken from [6,7,8,9,10] and QMC estimates are from [10,11,12].

| Lattice | $U/t$ | $t'/t$ | $N$ | $S$ | $\vec{K}$ | $\Gamma_{C_{4v}}$ | $S_m(\pi,\pi)$ SEMF | $S_m(\pi,\pi)$ Exact | $S_c(\pi,\pi)$ SEMF | $S_c(\pi,\pi)$ Exact |
|---|---|---|---|---|---|---|---|---|---|---|
| $4\times 4$ | 4 | 0 | 10 | 0 | (0,0) | $A_1$ | 0.732 | 0.73 | 0.509 | 0.506 |
| $4\times 4$ | 4 | 0 | 14 | 0 | (0,0) | $B_1$ | 2.198 | 2.14 | 0.428 | 0.4242 |
| $4\times 4$ | 8 | −0.3 | 14 | 0 | (0,0) | $A_1$ | 0.944 | 0.965 | 0.285 | 0.279 |
| $4\times 4$ | 4 | 0 | 16 | 0 | (0,0) | $A_1$ | 3.656 | 3.64 | 0.386 | 0.385 |
| $4\times 4$ | 8 | −0.3 | 16 | 0 | (0,0) | $A_1$ | 4.996 | 4.985 | 0.192 | 0.192 |
| $6\times 6$ | 4 | 0 | 36 | 0 | (0,0) | $B_1$ | 6.028 | 5.82(3) | 0.409 | 0.418(2) |
| $8\times 8$ | 4 | 0 | 64 | 0 | (0,0) | $A_1$ | 9.616 | 8.2(2) | 0.429 | 0.412(2) |

**TABLE II.** Magnetic and charge structure factors computed from the SEMF approach compared to exact values. Symbols and references are the same as in Table I.